\documentclass[showpacs,10pt,prl,twocolumn]{revtex4-1}
\usepackage{graphicx}
\usepackage{hyperref}
\usepackage{amsmath}
\usepackage{amssymb}
\usepackage{xcolor}
\newcommand{\bra}[1]{\left\langle#1\right|}
\newcommand{\ket}[1]{\left|#1\right\rangle}

\newcommand{\ba}{\begin{align}}
\newcommand{\ea}{\end{align}}
\newcommand{\dma}{\hat \rho_{Q}} %density matrix of subsystem A
\newcommand{\hop}{\hat{H}} %hamiltonian operator
\newcommand{\pop}{\hat p} %momentum operator
\newcommand{\xop}{\hat x} %position operator
\newcommand{\Xop}{\hat X} %position operator
\newcommand{\Sop}{\hat S} % operator S
\newcommand{\Uop}{\hat U} % operator U
\newcommand{\Udop}{\hat U^\dagger} % operator U dagger
\newcommand{\comm}[2]{\left[#1,#2\right]} %commutator
\newcommand{\Trm}{{\mathrm{T}}}

\def\XXint#1#2#3{{\setbox0=\hbox{$#1{#2#3}{\int}$}
     \vcenter{\hbox{$#2#3$}}\kern-.5\wd0}}

\begin{document}

\title{Quantum Lubricity}
\author{Tommaso Zanca$^{1}$, Franco Pellegrini$^{1,2}$, Giuseppe E. Santoro$^{1,2,3}$, Erio Tosatti$^{1,2,3}$}
\date{\today}

\affiliation{
$^1$ SISSA, Via Bonomea 265, I-34136 Trieste, Italy\\
$^2$ CNR-IOM Democritos National Simulation Center, Via Bonomea 265, I-34136 
Trieste, Italy\\
$^3$ International Centre for Theoretical Physics (ICTP), P.O.Box 586, I-34014 
Trieste, Italy
}

\begin{abstract}
{
The quantum motion of nuclei, generally ignored in sliding friction, can become important for an atom, ion, or light molecule sliding in an optical lattice. The density-matrix-calculated evolution of a quantum Prandtl-Tomlinson model, describing the frictional dragging by an external force of a quantum particle, shows that classical predictions can be very wrong. The strongest quantum effect occurs not for weak, but for strong periodic potentials, where barriers are high but energy levels in each well are discrete, and resonant tunnelling to excited states in the nearest well can preempt classical stick-slip with great efficiency. The resulting  permeation of otherwise impassable barriers is predicted to cause quantum lubricity.
}
\end{abstract}

\maketitle

Quantum effects in sliding friction, despite some early and laudable work~\cite{Mueser_PRL04,Mueser_JCP05,Volokitin_PRL11} have 
not been discussed very thoroughly so far. In most cases in fact the forced motion of atoms, molecules and 
solids is considered, and simulated, only classically.  The quantum effects that 
may arise at low temperatures, connected with either quantum freezing of the 
phonons or a slight quantum smearing of classical energy barriers, are not 
generally deemed to be dramatic and have received very little attention.
At the theoretical level in particular, quantum frictional phenomena were not pursued after and 
beyond those described by path-integral Monte Carlo in the 
commensurate Frenkel-Kontorova model~\cite{Mueser_PRL04,Mueser_JCP05}.
Possible reasons for this neglect are the scarcity of well defined frictional 
realizations where quantum effects might dominate and, on the theoretical side, 
the lack of easily implementable quantum dynamical simulation approaches. 

Cold atoms \cite{Bloch:rev} and ions \cite{Karpa_PRL13} in optical lattices offer brand new opportunities to explore 
the physics of sliding friction, including quantum aspects. Already at the 
classical level, and following theoretical suggestions~\cite{Benassi_Nat11}, 
recent experimental work on cold ion chains demonstrated important phenomena 
such as the Aubry transition~\cite{Bylinskii_Sci15,Bylinskii_Nat16}. 
The tunability of the perfectly periodic optical potential which controls the 
motion of atoms or ions should make it possible to access regimes where quantum 
frictional effects can emerge.
Here we show, anticipating experiment, that a first, massive quantum effect will 
appear already in the simplest sliding problem, that of a single particle forced 
by a spring $k$ to move in a periodic potential --- a quantum version of the 
renowned Prandtl-Tomlison model -- a prototypical system that should also be 
realisable experimentally by a cold atom or ion dragged by an optical tweezer.   
As we will show, the main quantum effect, amounting to a force-induced 
Landau-Zener (LZ) tunnelling, is striking because it shows up preferentially for 
strong optical potentials and high barriers, where classical friction is large, 
but resonant tunnelling to a nearby excited state can cause it to drop --- a 
phenomenon which we may call quantum lubricity.    

Our model consists of a single quantum particle of mass $M$ in the one-dimensional 
periodic potential created, for instance, by an optical lattice, of strength $U_0$ and 
lattice spacing $a$. The particle is set in motion by the action of a harmonic 
spring $k$, representing for instance an optical tweezer, which moves with 
constant velocity $v$:
%Eq. 1
\begin{equation}\label{eq:freeham}
\hop_{\rm Q}(t) = \frac{\pop^2}{2M} + U_0 \sin^2 \left( \frac{\pi}{a}\xop 
\right) + \frac{k}{2} \left( \xop-vt \right)^2 \;.
\end{equation}
The forced motion gives the particle an energy that, in a frictional steady 
state, is removed by dissipation in a thermostat. As pioneered by Feynman and Vernon~\cite{Vernon}, 
such a dissipation can be introduced by means of a harmonic bath~\cite{Weiss:book} 
%Eq. 2
\begin{equation}
\hop_{\rm int} = \sum_i \left( \frac{\pop_i^2}{2m_i}+\frac{1}{2} m_i \omega_i^2 
\Big( \xop_i - \frac{c_i}{m_i\omega_i^2} \Xop \Big)^2 \right)  \;,
\end{equation}
where each oscillator $\xop_i$ couples, through an interaction coefficient 
$c_i$, to the ``periodic position'' of the particle  
$\Xop=\sin \left( \frac{2\pi}{a} \xop \right)$. The coefficients $c_i$ 
determine the coupling strength of the bath, through the spectral function  
$J(\omega) = \hbar \sum_i \frac{c_i^2}{2 m_i \omega_i} \, \delta(\omega-\omega_i)$, 
which we choose of the standard Caldeira-Leggett ohmic form 
$J(\omega)=2 \alpha \hbar^2 \omega e^{-\omega/\omega_c}$, 
where $\omega_c$ sets the high-energy cutoff. 

% Fig. 1
\begin{figure}
\centering
\includegraphics[width=8cm]{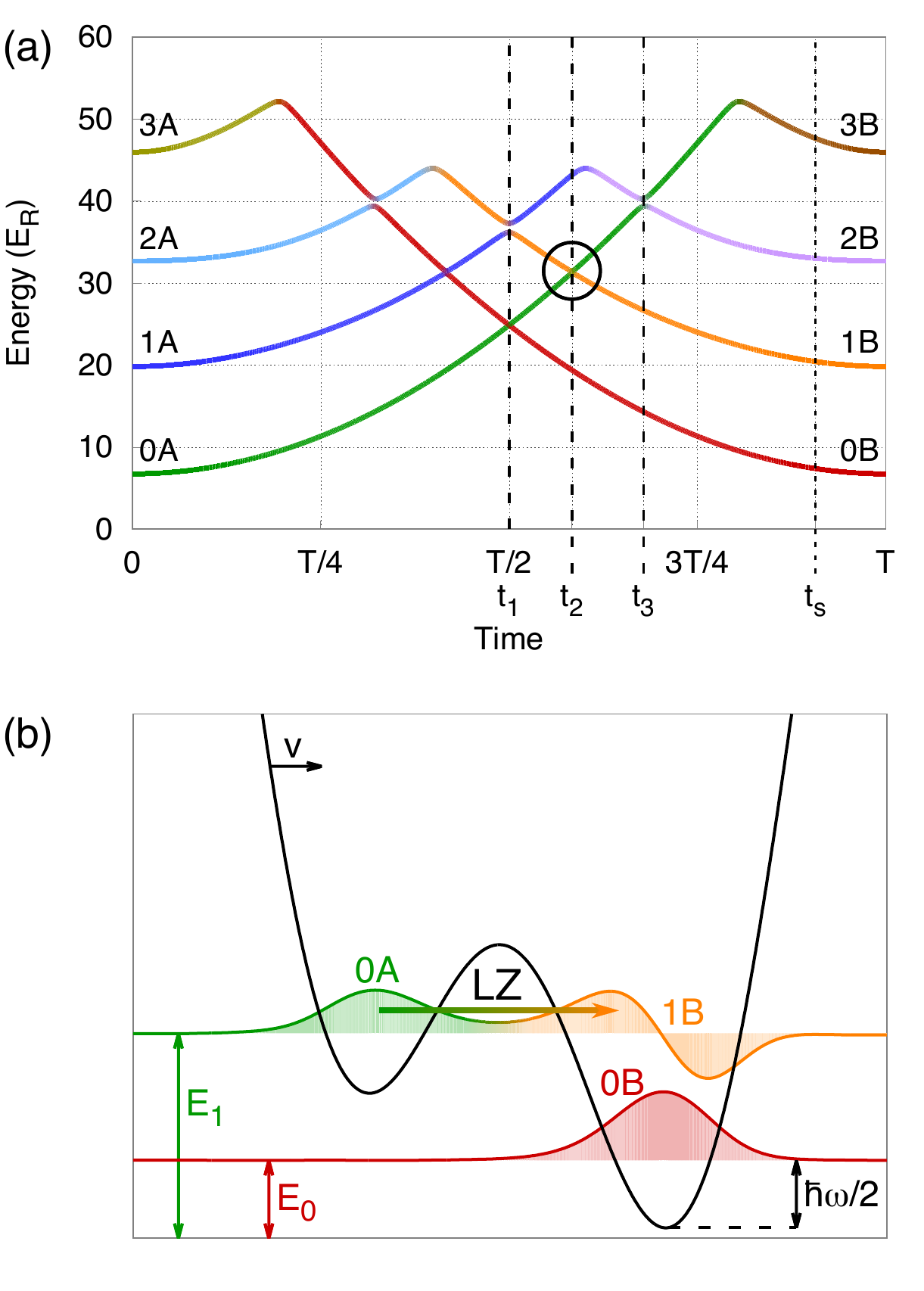}
\caption{ (Color online) 
(a) The four lowest instantaneous eigenvalues of a particle that is adiabatically driven by 
the harmonic trap from a periodic potential minimum to the nearest one. 
Note the avoided-crossing gaps associated with tunnelling events encountered during the dynamics at times $t_1$, $t_2$ and $t_3$. 
The circle highlights the resonant tunnelling described in text and represented in (b).
(b) A pictorial sketch of the tunnelling event in which a particle in the ground level 
of the left well (A) resonantly tunnels into the first excited level of the right well (B). }
\label{fig:Fig1}
\end{figure}

We can understand the basic mechanism leading to quantum frictional dissipation 
by considering the instantaneous eigenstates of $\hop_{\rm Q}(t)$, shown in Fig.~\ref{fig:Fig1}(a) for a reduced Hilbert space with 4 states per well.
Denote by $\Trm=a/v$ the time period in which the driving spring moves by one lattice spacing:
at $t=0$, when the harmonic potential is centered at $x=0$, the lowest eigenstate is essentially 
coincident with the lowest Wannier state in the $x=0$ potential well. 
As the harmonic spring moves forward, at $t=t_1=\Trm/2$, the particle negotiates the perfect double-well state 
between $x=0$ and $x=a$, where all pairs of left and right levels anticross. 
The LZ  ``diabatic'' transition rate (population of the excited state after the anticrossing) between levels $E_{n} (t)$ and $E_{n'}(t)$ is  
%Eq. 3
\begin{equation}\label{eq:LZ}
P_{n\to n'}= e^{-\frac{\pi \Delta_{nn'}^2}{2\hbar v \alpha_{nn'}}} = e^{-\frac{v_{n\to n'}}{v}} 
\end{equation}
where $\alpha_{nn'}$ is the relative slope of the two eigenvalues involved, $E_n$ and $E_{n'}$,   
$\Delta_{nn'}$ their anticrossing gap, and $v$ is the speed.

At the anticrossing at $t_1=\Trm/2$ between ground states at $x=0$ and $x=a$,
due to the large barrier the states are very localized and the gap, here $\Delta_{01}$, is exceedingly small.
For very small velocity,
nonetheless, 
 $v \ll  v_{0\to 1} = \pi \Delta_{01}^2 /[2 \hbar \partial_x |(E_1-E_0)|)$,
the LZ transition rate $P_{0\to 1}$~\eqref{eq:LZ}, which as we shall see is 
proportional to the frictional dissipation is negligible. 
In that low velocity case, a quantum particle is transmitted adiabatically 
without friction. This is therefore a regime, which one might designate of 
{\it quantum superlubricity}, where friction vanishes non analytically as in 
Eq.~\eqref{eq:LZ} in the limit of zero speed -- totally unlike the classical case, where 
friction vanishes linearly with $v$ (viscous friction). 
Quantum superlubricity should be realized at sufficiently low temperature, to be thermally destroyed in favor of viscous lubricity as soon 
as temperature $T$ is large enough to upset the LZ physics behind the mechanism. 
This, however, is not expected to occur until $T$ becomes considerably larger than the tunnelling gap $\Delta_{01}$, 
as a recent study on the dissipative LZ problem has confirmed \cite{Arceci_PRB17}.  

Moving on to larger speeds $v \gg v_{0\to 1}$, the particle, unable to negotiate tunnelling adiabatically, 
remains diabatically trapped with large probability $P_{0 \to 1}$ in the lowest $0A$ Wannier state even for $t>\Trm/2$.  
Only at a later time, $t=t_2$, the rising level becomes resonant with the first excited state {1B} of the $x=a$ well. 
As this second gap $\Delta_{12}$ is now much larger, the LZ diabatic rate drops and the particle transfers with large adiabatic 
probability from the $A$ to the $B$ well for driving speeds $v_{0\to1}\ll v\ll v_{1\to2}$. 
Once the first excited  $1B$ state in the $x=a$ well is occupied, the bath exponentially sucks out the 
excess energy and thermalizes the particle to lowest $0B$ level. 
That amounts to dissipation which is paid for by frictional work done by the external force. 
The $0A\to 1B$ quantum slip between neighboring wells preempts by far the 
classical slip, which would take place when the rising classical minimum disappears, at  
$t_s = (\pi U_0 / k v a) \sqrt{1-(ka^2/2 \pi^2 U_0)^2} +(a/2\pi v) \cos^{-1}(-ka^2/2\pi^2 U_0) > t_{2}$ . 

To calculate the frictional dissipation rate, we describe the particle motion by 
means of a weak coupling Born-Markov quantum master equation (QME), based on a 
time-evolving density matrix (DM) $\dma(t)$~\cite{Yamaguchi_PRE17,Arceci_PRB17}, 
whose equation of motion is
%Eq 4
\begin{equation}
\frac{d}{dt}\dma(t)=\frac{1}{i\hbar}\comm{\hop_X(t)}{\dma(t)} - \left( 
\comm{\Xop}{\Sop(t)\dma(t)} +\text{H.c.}\right)\;,
\end{equation}
\\
where $\hop_X(t) = \hop_Q(t) + 2\hbar\alpha\omega_c \Xop^2$. 
The operator $\Sop(t)$, which is in principle \cite{Yamaguchi_PRE17} a bath-convoluted $\Xop$ given by
$\Sop(t) = \frac{1}{\hbar^2} \int_0^t d\tau\,C(\tau)\,\Uop_X(t,t-\tau)\,\Xop\,\Udop_X(t,t-\tau)$, 
will be approximated, in the basis of the instantaneous eigenstates $\ket{\psi_k(t)}$ of the system Hamiltonian $\hop_{X}(t)$  
as $\Sop(t)=\sum_{k,k'} S_{k,k'}(t)  \ket{\psi_k(t)}\bra{\psi_{k'}(t)}$ with
%Eq. 5
\begin{equation}
S_{k,k'}(t) \approx \frac{1}{\hbar^2} \bra{\psi_k(t)} \Xop \ket{\psi_{k'}(t)} \, \Gamma(E_{k'}(t)-E_k(t)) \;,
\end{equation}
where $\Gamma(E^+)\equiv\int_0^{+\infty} d\tau\,C(\tau)\,e^{i\left(E+i 0^+\right)\tau/\hbar}$ 
is the rate for a bath-induced transition at energy $E$, and $E_k(t)$ is the instantaneous eigenvalue associated to $\ket{\psi_k(t)}$.
Recent work on the dissipative LZ problem \cite{Arceci_PRB17} has shown that this approximation is perfectly safe, 
when the coupling to the bath is weak, in an extended regime of driving velocities $v$.  
The QME is then solved in the basis of the Wannier orbitals of the unperturbed particle in the periodic potential.

%Fig. 2
\begin{figure}
\centering
\includegraphics[width=8cm]{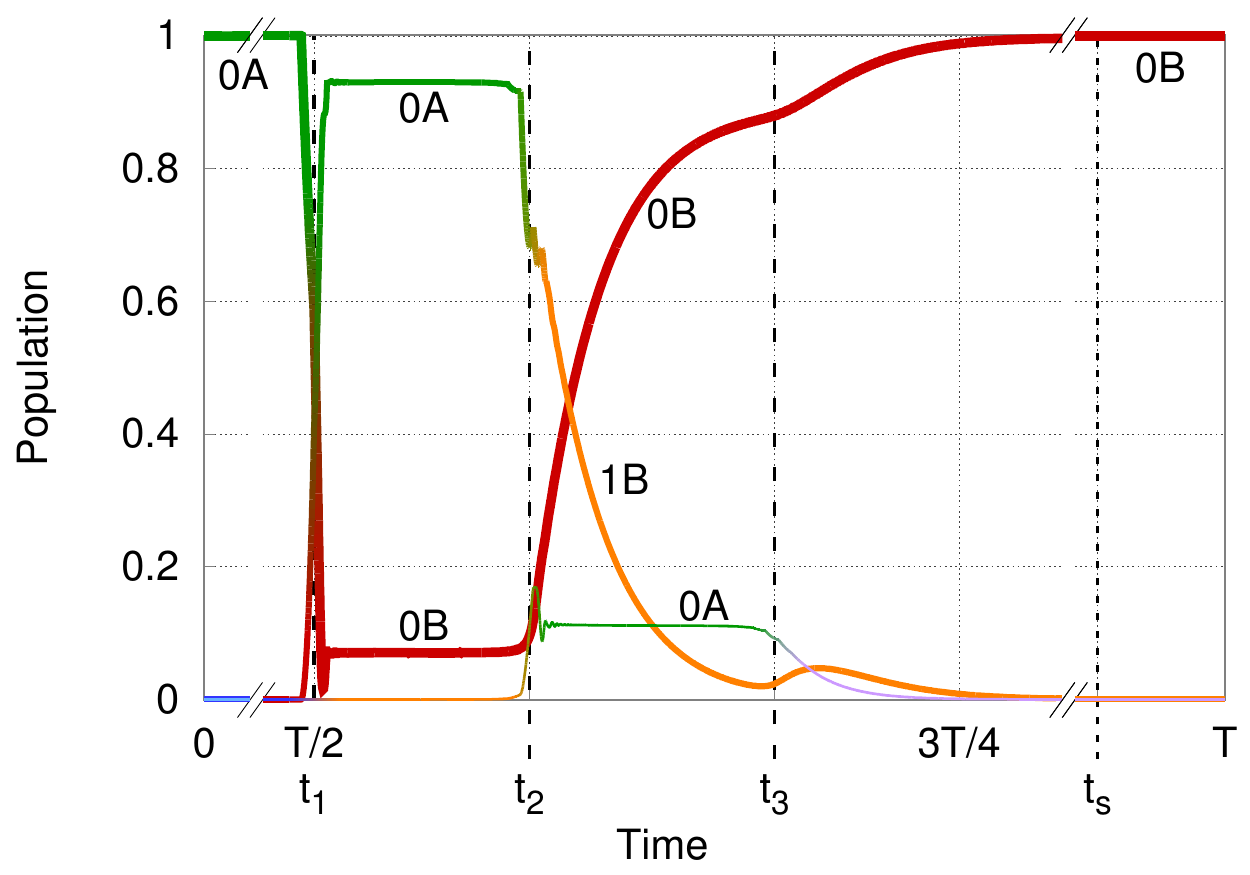}
\caption{(Color online) Time-dependent population of instantaneous eigenstates for $v=4\cdot10^{-4}\,E_Ra/\hbar$, $U_0=38.5\,E_R$, $k=190\,E_R/a^2$,
where $E_R=\pi^2\hbar^2/(2Ma^2)$ is the recoil energy, corresponding to the double-well potential configuration sketched in Fig.~\ref{fig:Fig1}(b).  
Lines of decreasing thickness are used for higher eigenstates. 
The ohmic coupling strength is here $\alpha=0.002$ with a cutoff $\omega_c=12\, E_R/\hbar$, and temperature $T=1\, E_R/k_B$.
}
\label{fig:population}
\end{figure}
%....................
%
Figure \ref{fig:population} shows, for an arbitrary but reasonable choice of 
parameters, the time-dependent population probability of the first three 
instantaneous eigenstates, $P_k(t)=\langle \psi_k(t) | \dma | \psi_k(t) \rangle$, 
over one period of forced particle motion in the $v_{0\to1}\ll v\ll v_{1\to2}$ regime. 
As qualitatively sketched, despite the slow motion the probability of 
the $0A\to0B$ {adiabatic transition} to the right well ground state 
at $t_1=\Trm/2$ is already very small, and LZ dominates this first level crossing 
keeping diabatically the particle in the left $A$ well. 
At the second $1\to2$ crossing where the gap $\Delta_{12}$ is much larger, $P_{1\to2}$ is suppressed, 
and the $1^{\rm st}$ excited level of the right well ($1B$) becomes strongly populated. 
Following that, the bath exponentially relaxes $P_k(t)$ down to the right well ground state.

\begin{figure}
\centering
\includegraphics[width=8cm]{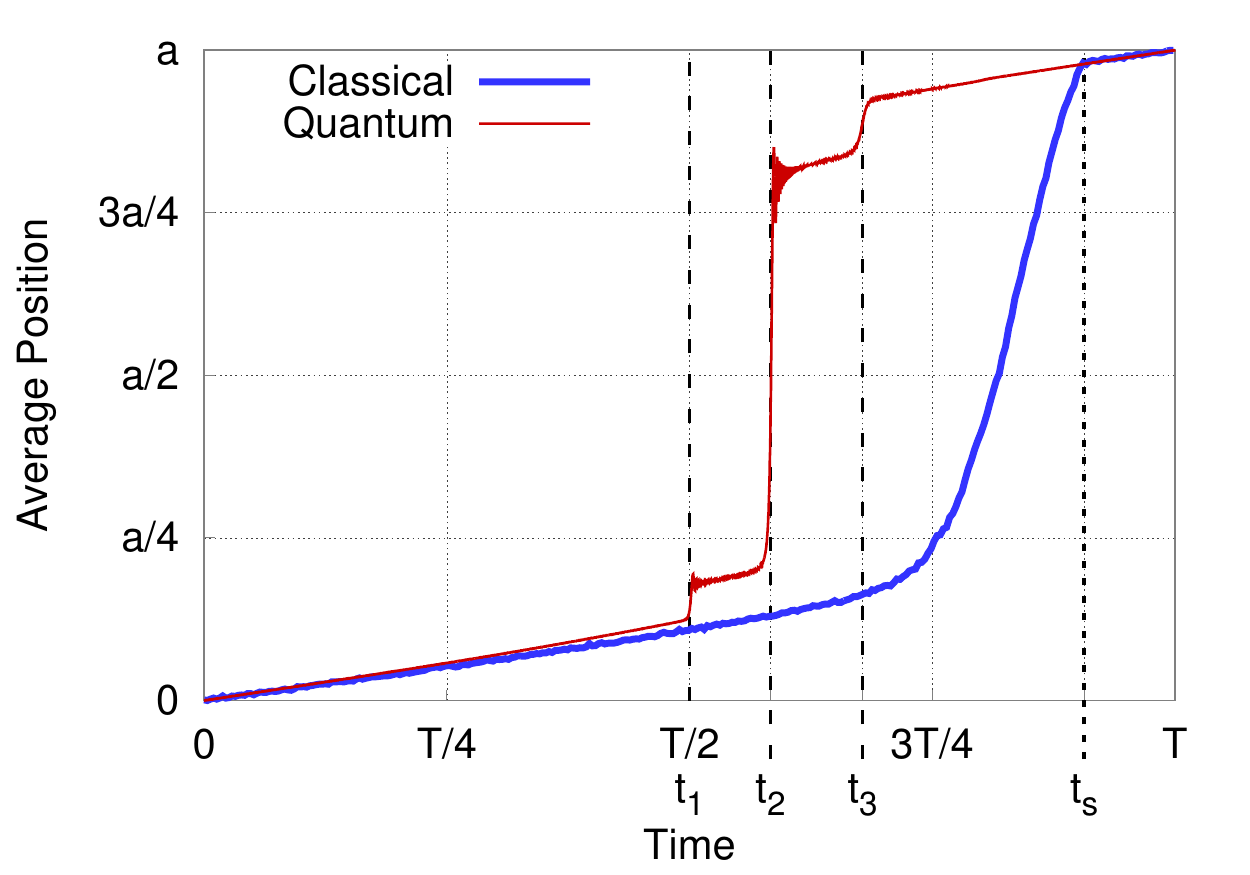}
\caption{(Color online)
Average position of the particle versus time, in the quantum and classical cases.  
Parameters are identical to those of Fig.~\ref{fig:population}. 
Most of the ``slip'' of the quantum particle goes through the excited-state resonant tunnelling, 
taking place at $t_{2}$ beyond the symmetric moment $t_1=\Trm/2$ between the two potential wells.  
The dashed line shows the position of the classical ``spinodal'' moment $t_s$, where the $x=0$ 
local potential minimum disappears and the particle is forced to slip.
}
\label{fig:position}
\end{figure}

The mechanism just described predicts an advancement of the average position of 
the particle, as well as a corresponding onset of dissipated power, very different
from those of ordinary Langevin frictional dynamics~\cite{Weiss:book}, which, 
with all parameters except $\hbar$ the same as in the quantum case, describes 
the classical forced sliding of the same particle.
Figure \ref{fig:position} compares the average particle position versus time in 
the quantum and classical cases.  The ``quantum slips'' occur rather suddenly, 
reflecting the abruptness of level crossing events and connected barrier 
passage. In particular, the main quantum slip occurs, for the parameters used in 
Fig.~\ref{fig:position}, precisely when the instantaneous Wannier ground level the left well 
is resonantly aligned with the first excited Wannier level in the neighboring well. 

%Fig. 3

\begin{figure}
\centering
\includegraphics[width=8cm]{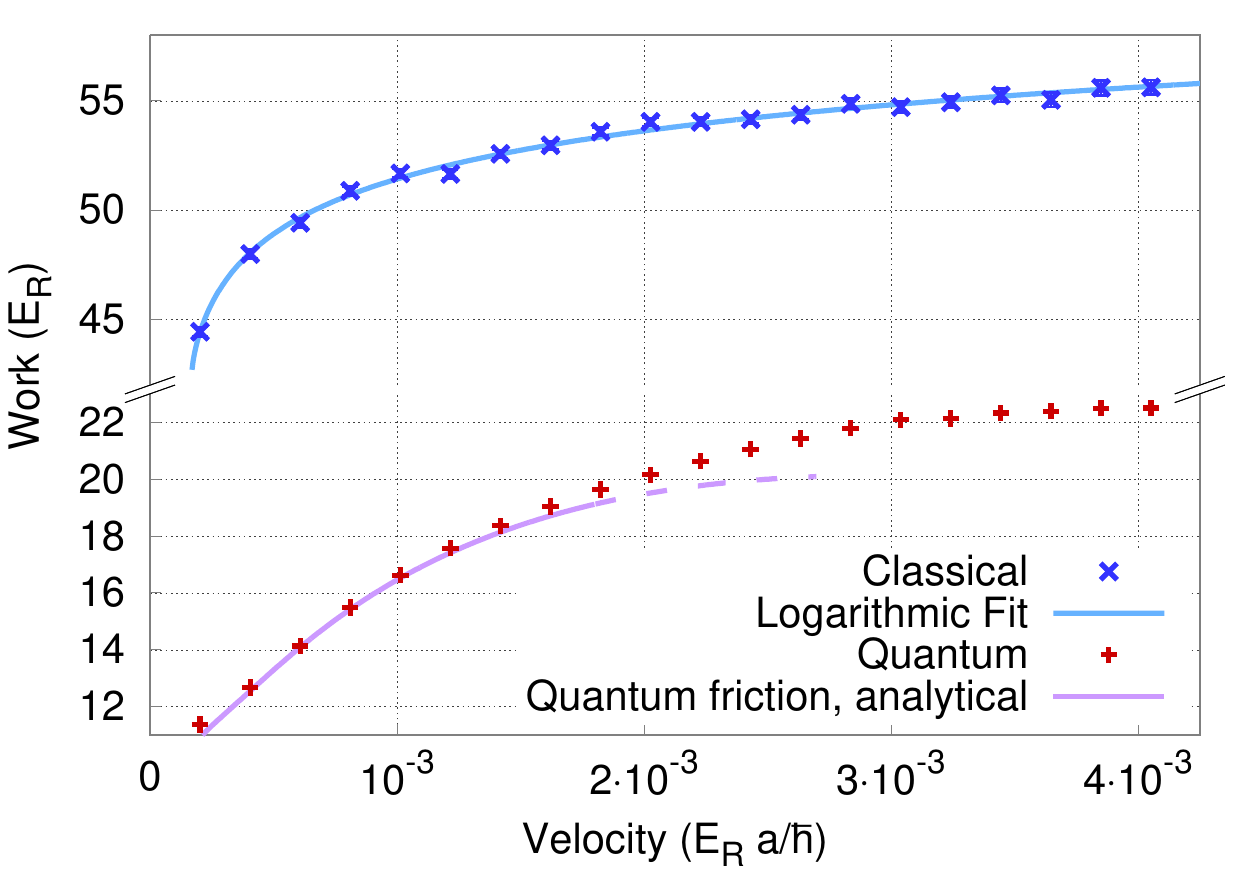}
\caption{(Color online) 
Frictional dissipation rate for classical and quantum sliding vs driving velocity.
Note the large reduction of dissipation induced by the resonant quantum tunneling: quantum lubricity. 
Parameters are identical to those of Fig.~\ref{fig:population}.}
\label{fig:work}
\end{figure}

Because it occurs at a lower spring loading, the resonant barrier permeation 
strongly reduces the overall mechanical friction work exerted by the pulling spring. 
Figure \ref{fig:work} shows the amount of energy absorbed by the bath (friction) 
at the end of each period as a function of velocity.
In the classical case the friction grows logarithmically with speed, due to thermally activated slip, 
as is well known for stick-slip at finite temperature \cite{Gnecco00, Sang_PRL01,Dudko_CPL02,Vanossi_RMP13} 
%Eq. 6
\begin{equation}
W_{cl}=a+b\,\ln^{2/3}\left(c \,v\right).
\end{equation}
with  constants $a=42.5\,E_R$, $b=6.11\,E_R$ and $c=5.92\cdot10^3\,\hbar/E_R a$ providing the best fit in our case.
The quantum dissipation rate is by comparison smaller by a factor $\sim 3$. 
It is well approximated through the Landau-Zener probabilities Eq.~\eqref{eq:LZ}
of transition from the $n^{\rm th}$ to the $(n+1)^{\rm th}$ eigenstate:
%Eq. 7
\begin{align}
%W_q(v) &= P_{0\to 1}(v) \left[ (1-P_{1\to 2}(v)) (E_1-E_0) + \cdots \right] \;.\\
W_q(v)&\approx P_{0\to1}(v)\left[ (1-P_{1\to 2}(v)) \left( E_1-E_0\right) +\right.\nonumber\\
&\quad \left. P_{1\to 2}(v)\left( 1-P_{2\to 3}(v) \right)\left( E_2-E_1 \right) \right]\,,
\label{eq:landau-zener}
\end{align}
with $\Delta_{01}=5.19\cdot10^{-2}\,E_R$, $\Delta_{12}=3.03\cdot10^{-1}\,E_R$, $\Delta_{23}=8.83\cdot10^{-1}\,E_R$; $\alpha_{01}=1.43\cdot10^2\,E_R/a$, $\alpha_{12}=1.38\cdot10^2\,E_R/a$, $\alpha_{23}=1.46\cdot10^2\,E_R/a$;
$v_{0\to1}=2.96\cdot10^{-5}\,E_Ra/\hbar$, $v_{1\to2}=1.05\cdot10^{-3}\,E_Ra/\hbar$, $v_{2\to3}=8.40\cdot10^{-3}\,E_Ra/\hbar$.
Dissipation requires in fact, to start with, that the system does not LZ tunnel, 
so that $P_{0\to1}>0$. The amount of power absorbed by the bath equals  
the probability to populate the first and higher excited states times their energy 
difference with the ground state.
Eq.~(\ref{eq:landau-zener}) is approximate first of all because it does not include higher excited states.
Moreover, it is only valid when velocity is low enough that the cooling 
rate $\gamma_c \gg v/a$, and the particle loses all its kinetic energy before 
encountering the subsequent slip, which is not satisfied for the larger velocities.
%The dots in Eq.~(\ref{eq:landau-zener}) 
%include contributions from higher excited states.
It is clear that, unless temperature is too high, quantum tunnelling through 
the barrier always preempts classical negotiation of the barrier, causing 
friction to be necessarily smaller than classical friction. In this sense we 
can speak of {\it{quantum lubricity}}.

Despite its conceptual simplicity, this form of quantum lubricity has not been addressed 
experimentally but should be well within experimental reach for cold atoms/ions in optical lattices. 
The parameters used in our simulations assume a particle with the mass $M$ of 
$^{171}$Yb, and a lattice spacing $a=500$ nm.
The lattice potential is taken to be $U_0=38.5\, E_R$, in terms of the recoil energy $E_R=\pi^2\hbar^2/(2Ma^2)$. 
The corrugation parameter $\eta=(\omega_l/\omega_0)^2$, defined \cite{Bylinskii_Sci15} as the confinement ratio 
of the lattice intra-well vibrational frequency $\omega_l= 2\sqrt{U_0\,E_R}/\hbar$
to the harmonic trap (the optical tweezer pulling spring) vibrational frequency $\omega_0=a\sqrt{2k E_R}/\pi\hbar$,
is set equal to $\eta=4$, so that the overall potential energy has just two minima.
This automatically sets the value of the optical tweezer spring constant at $k=190\, E_R/a^2$. 
Finally, the assumption of weakly coupled Ohmic environment, with $\alpha=0.002$ and $\omega_c=12\, E_R/\hbar$, 
necessary for a consistent perturbative theory, can be realized by a judicious choice of cooling strengths.
The values adopted for $\alpha$ and $\omega_c$ correspond to a cooling rate $\gamma_c\approx 0.018\,E_R/\hbar$.
In order to make the bath effective during the dynamics, the condition on the optical tweezer velocity 
$v<\gamma_c\,a$ must be satisfied, leading to a time-scale of the optical tweezer motion much larger than the 
period of vibrations in the lattice well: $v/a\ll\omega_l$.

In summary, comparison of classical and quantum stick-slip friction for a particle sliding in a periodic potential reveals major differences. 
A classical particle slides from a potential well to the next by overcoming the full potential barrier.  
A quantum particle can permeate the barrier by resonant tunnelling to an excited 
state, a process suddenly and narrowly available at a well defined position of 
the harmonic trap, leading to discontinuous transfer to the next well, as shown 
in Fig.~\ref{fig:position}. This quantum slip preempts the classical slip, 
giving rise to quantum lubricity. The potential energy accumulated by the 
particle during sticking, and frictionally dissipated at the quantum slip, is 
just the amount sufficient to reach the resonant condition with the excited 
state in the next well. Conversely, the classical potential energy increases 
necessary for classical slip is close to the top of the barrier, with a 
correspondingly larger amount of dissipated energy during and after the slip.
In addition to this quantum lubricity effect, a regime of quantum superlubricity is in principle expected at sufficiently low temperatures, 
where the friction growth with velocity should begin non-analytically, with all derivatives vanishing.  
The natural extension of these predictions to many-particle system will be of interest in the future. 

Research supported by the EU FP7 under ERC-MODPHYSFRICT, Grant Agreement No. 
320796, and in part by COST Action MP1303.

\bibliography{QuantumLubricityFinal}

\end{document}